\documentclass{sigkddExp}
\usepackage{adjustbox}
\usepackage{booktabs}

\begin{document}
%

\title{Business Policy Experiments using Fractional Factorial Designs: Consumer Retention on DoorDash}
%

\numberofauthors{3}
%


\author{
%
\alignauthor Yixin Tang \\
       \affaddr{DoorDash, Inc.}\\
        \affaddr{303 2nd San Francisco, CA, USA}\\
       \email{yixin@doordash.com}
\alignauthor Yicong Lin\\
       \affaddr{DoorDash, Inc.}\\
        \affaddr{303 2nd San Francisco, CA, USA}\\
       \email{nicole.lin@doordash.com }
\alignauthor Navdeep S. Sahni\\
       \affaddr{Stanford GSB}\\
       \affaddr{655 Knight Way, Stanford, CA, USA}\\
       \email{navdeep.sahni@stanford.edu }
}

\maketitle
\begin{abstract}
This paper investigates an approach to both speed up business decision-making and lower the cost of learning through experimentation by factorizing business policies and employing fractional factorial experimental designs for their evaluation. We illustrate how this method integrates with advances in the estimation of heterogeneous treatment effects, elaborating on its advantages and foundational assumptions. We empirically demonstrate the implementation and benefits of our approach and assess its validity in evaluating consumer promotion policies at DoorDash, which is one of the largest delivery platforms in the US. Our approach discovers a policy with 5\% incremental profit at 67\% lower implementation cost. 
\end{abstract}

\section{Introduction}
\hfill \break

Businesses commonly employ randomized experiments to select the optimal policy among several options. This approach involves evaluating outcomes across randomly chosen units exposed to experimental policies to infer the most effective policy. While the simplicity of this method is appealing, it can also be tedious and delay the policy evaluation process, and thus the business innovation process. This is because setting up counterfactual policies concurrently entails setup costs that increase with the number of test policies, and it can take time to gather the necessary sample size to obtain statistically conclusive results.\footnote{The literature has discussed several other aspects of this strategy including designing experiments for for decision-making \cite{feitBerman2019}, estimating long-term effects \cite{athey2019surrogate}\cite{yang2023targeting}, presence of network effects \cite{eckles2016estimating}, and parallel experimentation \cite{waisman2019parallel}.} \\

This paper investigates an approach to solve these challenges by considering the factorization of business policies and using factorial experimental designs to evaluate policies. Using a model we show how this approach can be combined with advances in the estimation of heterogeneous treatment effects, and discuss its benefits and its underlying assumptions. We empirically demonstrate our approach and its benefits and assess its validity in evaluating consumer promotion policies at DoorDash, which is one of the largest delivery platforms in the US. \\

Our motivation comes from two observations. \\

First, since business policies touch upon several different functions of the business, testing them involves significant setup costs, which distinguishes experimenting with them from typical web experiments \cite{kohavi2009controlled}. Consider a simple scenario in which a digital retail platform is considering spreading out the promotional incentives it gives to consumers over time, as opposed to providing them upfront which is the status quo. Implementing the test scenario realistically requires engineering several aspects of the platform, including but not limited to random assignment of the users into experimental buckets; generating separate promo codes for the test and control policies; making the users-promo codes mapping visible to the systems that use them (e.g., CRM, platform UI, Analytics); designing relevant customer-facing messages (emails, promo banners shown on the platform home page); designing matching user browsing experience (platform ranking of options may change accordingly); specifying the matching call center customer experience and training the sales support team accordingly. Merely identifying such touchpoints takes effort. Then each of these aspects needs to be configured correctly, coordinated across different teams owning them, and tested thoroughly before the experiment starts. A bug in the workflow can deter the user experience, harm the business, and render the test useless. \\

Further, note that the effort required in such implementations increases with scale of the number of variants, which increases more if offline experience change is also involved.  Indeed the literature has documented several instances in which complex experiment designs led to mistakes in implementation \cite{sahni2019experimental,miller2022sophisticated}. \\

Second, business policies are often regarded as monolithic units and not as combinations of distinct components at the testing stage. Consequently, businesses often implement separate policies and compare them via A/B testing \cite{kohavi2017surprising}. This approach, while tedious, can work satisfactorily in the early design stages when the policy space is relatively unexplored and the magnitude of improvements (effect sizes) is large.  However, the statistical power of experimentation can be a constraint challenging the ability to identify improvements in later stages when significant advancements have been made and the experiments' effect sizes tend to be lower. Also, given the rate of true improvement becomes increasingly lower, companies that simply follow a 0.05 $p$-value shipping criterion will have a higher false positive risk \cite{inproceedings}.\footnote{For example, when the rate of actual improvement is $5\%$ the false positive rate can be as high as $37\%$.}  This can lead to incorrect learning and unnecessary production costs. \\

In this paper, we use an analytical model to show the impact of factorization of policies, and how one can use fractional factorial designs to test them. We specify the underlying assumption it makes, characterize the value of factorization relative to other approaches, and integrate it with advances in estimation of heterogeneous treatment effects. \\

Next, we implement this method at DoorDash to demonstrate its usage, assess its validity and quantify its value. In our context, we take on the platform's problem of re-engaging inactive users using promotions. Our objective is to find the promo configuration that will have the maximum impact, holding constant the monetary cost of giving the promo. In other words, given the company's willingness to spend per customer retention, what is the best structure of the promo?  How does the optimal promo structure vary across customers? Our setting is typical because experimenting with promotional policies is tedious and costly. \\

Following the above approach, we factorized retention policies into four factors and nine levels in total, which is a $2\times2\times2\times3$ design. In our implementation, we use a fractional factorial orthogonal experiment design with eight experimental arms. Using this experiment, we are able to estimate the effect of each of the 24 possible policies, assuming no interaction between the factor's effects. Given the availability of individual-level covariates, we are able to estimate heterogeneous effects which enable us to recommend optimal personalized policies. \\

Additionally, to test the modeling assumption of no interactions, we launch an additional ``held out'' experimental arm that is held out in estimation and used only for the purpose of evaluation. \\

In our application we are able to evaluate personalized effects of 24 policies at the cost of implementing an experiment with 8 policies, showcasing a 67\% reduction in implementation cost. Given the orthogonality of the assigned factors, we gain statistical efficiency because each factor's impact is estimated by leveraging the whole sample. Overall, we are able to recover a policy that increases profits by 5\%.\\

To test our model assumptions we predict the impact of the held-out experimental arm relative to the other eight arms and compare this prediction to the actual difference. We repeat this procedure separately for different individual groups. We detect no significant differences between the model prediction and the actual differences in these cases, showing that our data support the modeling assumptions. \\

Overall, this paper shows a practical and rigorous way of speeding up business decision making by factorizing the business policy space. As a consequence of our approach, the company might reduce the number of conducted experiments but increase the amount of learning per experiment because the learning is at the factor level and not at the policy level. This paper draws on the extensive body of research on multivariate testing in statistics \cite{box2005statistics} to combine the statistical efficiency of fractional factorial designs with modern methods for estimating heterogeneous treatment effects, and illustrating its benefit in testing business policies. It is also related to Conjoint Analysis work in Marketing \cite{green1978conjoint,green1990conjoint} which typically involves evaluating products by characterizing them as bundles of attributes. Our approach is inspired by this literature and extends the approach to settings such as digital platforms where cross-sectional user behavioral data is available (as opposed to a smaller longitudinal survey data in Conjoint), and some of the assumptions made to speed up experimentation can be tested. While the most commonly used Conjoint Analysis designs ask respondents to compare profiles, our policy experiments expose one profile to each user. Therefore, although the objective of estimating marginal effects and heterogeneous treatment effects is similar to Conjoint Analysis \cite{nicolehte}, our methodology is different.  Our approach is also related to the more recent literature on the estimation of the heterogeneity of treatment effects \cite{wager2018estimation,hitsch2018heterogeneous} as we apply that approach to multivariate testing. More generally, our approach is directionally related to structural econometric methods that make assumptions for a more efficient policy evaluation; the difference is that assumptions in structural models are more tightly grounded in economic and consumer theories \cite{reiss2007structural,chintagunta2018structural}. \\

The remainder of the paper is organized as follows. In Section 2, we break down the framework step by step and discuss its advantages compared to some other commonly used experiment methodologies. In Section 3, we provide the business context of this paper and discuss challenges that are not solvable by traditional A/B tests. Then we describe the key steps to applying this framework in this business context. In Section 4, we give an overview of the proposed framework and how one can apply it step by step. In Section 5, we discuss the design of the experiment, including campaign factorization, and some practical considerations about sample size calculation. In Section 6, we present our results, including end-to-end framework validation with out-of-sample variants, test the existence of heterogeneity, and discuss the benefits of this framework in terms of experimentation velocity, business impact, and how it creates unique opportunity for optimization using the HTE model. Section 7 concludes this paper by summarizing the challenges and solutions proposed in this paper.

\section{Framework}
\hfill \break

\subsection{Break down the policy space into factors}
\hfill \break

Policy factorization is one of the most important steps which sets a foundation for reducing the experimental space and concurrently testing different policies in later steps. First, we will discuss what constraints the selected factors need to satisfy and how to validate the model. We will also cover how we practically select the factors later in sections \ref{sec:businessCont} and \ref{sec:expDesign}. \\

Here, we introduce some terminology and notation used throughout the paper.

\begin{itemize}
  \item Variant: One variant is a unique candidate business policy
  \item Factor: One factor is a feature or attribute of a policy
  \item Level: Value of a factor. A factor often has two or more levels
  \item $n$: Total number of potential variants
  \item $I$: Total number of experimentation units
  \item $i$: Unit of experimentation; $i=1,2,...,I$
  \item $Y_i$: The primary business metric of unit $i$
  \item $F$: Total number of factors
  \item $L_f, f=1,2,3...,F$: The number of levels of the factor $f$
  \item $W_{ifl}, f=1,2,3...F, l=1,2,3...L_f$: The binary variable denoting if the unit $i$ 's factor $f$ is on level $l$
\end{itemize}

Given the notation, we can express the total number of potential variants as the product of the number of levels of each factor,

\begin{equation}n = \prod_{l=1}^{F} L_f\end{equation}

Our framework allows us to write $Y_i$ as a function of factors and levels. For example, a linear function looks like below

\begin{align} 
Y_i &= \sum_{f=1}^{F}\sum_{l=1}^{L_f} W_{ifl} \beta_{fl} \notag \\
&+ \sum_{f_1=1}^{F}\sum_{l_1=1}^{L_{f_1}} \left( \sum_{f_2=1,f_2\ne f_1}^{F}\sum_{l_2=1}^{L_{f_2}} W_{if_1l_1} \times W_{if_2l_2} \beta_{f_1l_1,f_2l_2} \right) \notag \\
&+ \sum_{f_1=1}^{F}\sum_{l_1=1}^{L_{f_1}} \left( \sum_{f_2=1,f_2\ne f_1}^{F}\sum_{l_2=1}^{L_{f_2}} \left( \sum_{f_3=1,f_3\ne f_1, f_2}^{F}\sum_{l_3=1}^{L_{f_3}} \right. \right.\\
&\left.\left.  W_{if_1l_1} \times W_{if_2l_2} \times W_{if_3l_3} \beta_{f_1l_1,f_2l_2,f_3l_3} \vphantom{\int_1^2} \right) \right) \notag \\
& ... \notag \\
&+ \epsilon_i \label{eq:fullmodel} 
\end{align}

where $i = 1,2,3..., I$, $I$ is the total number of units(users), $\epsilon_i$ is the error independent of $W_{ijk}$. Note that this model allows every combination of factor levels to have its own unique impact on $Y$. Consequently, the number of unknowns ($\beta$s) here will be the same as the number of the total number of potential variants which is $n$, so to estimate this ``full'' model we will need an experiment that implements all potential variants. 
\\

This framework allows us to make systematic assumptions about the interaction among factors that appear in lines two and three in equation \eqref{eq:fullmodel}. Given our purpose to simplify the implementation of the experiment, we make assumptions about the interaction terms and assume the factors to be additive with no interactions, that is,

\begin{equation} Y_i = \sum_{f=1}^{F}\sum_{l=1}^{L_f} W_{ifl} \beta_{fl} + \epsilon_i \quad (i = 1,2,3... I) \label{eq:ourmodel}\end{equation}

\subsection{Factorial Design}
\hfill \break

After selecting the factors, we can encode policies as unique combinations of factor levels. We can launch a full factorial experiment to measure the treatment effect of each combination of factor values, as in equation \eqref{eq:fullmodel}. However, in some instances, this can take unrealistic effort for the business to prepare all policies, for example in the context of testing customer promotions we consider in the empirical section. \\

We are hence interested in fractional factorial designs that reduce the number of the policies needed to be implemented in the experiment while still being able to estimate the treatment effect of all potential policies. There are multiple fractional factorial designs we can choose from, giving us the flexibility to trade off between the number of implemented policies and the estimated interaction between the factors.  \\

In cases where we assume no interaction effect among factors as in the model \eqref{eq:ourmodel}, all we need to do is identify the main effects. In such cases, we are good with a Resolution III experiment design \cite{box2005statistics} that aliases the main effects with two-factor interactions, i.e. the main effects are indistinguishable from the effects of two-factor interactions. \\

In cases where we expect significant interaction effects,  we would choose a higher resolution design, with a goal of not aliasing effects which we think are important with effects of less importance. We might end up with more runs when there exist higher-order interactions. Although using a lower resolution design yields a lower number of experimentally implemented variants, it is important to keep in mind the potential risk brought by aliasing. In the example used in this paper, we assume that there is no interaction effects between factors. To validate this, we launch another new experiment at the end to test the validity of the framework end-to-end.

\subsubsection{Other Fractional Factorial Designs}
\hfill \break

Fractional factorial designs \cite{factorial_book} are commonly used in applied statistics (e.g., Conjoint Analysis) as a way to reduce the number of variants that need to be tested, especially when the number of factors is large. One common method is to select one block from one of the single-replicate designs as the fraction to be used. For example, using $1/2^q$ fraction of a $2^p$ experiment gives us a $2^{p-q}$ fractional factorial experiment. Another common method is to start with an orthogonal array, and then use the array and its labeling to determine the defining relation and design. Sometimes, when the total number of combinations is not a power of $2$, we can choose the design that ensures the orthogonality,  for example, the Plackett-Burman design \cite{10.1093/biomet/33.4.305} can be used when the number of combinations is a multiple of $4$. There is also a group of designs, known as robust designs or parameter designs, which involve both noise factors and design factors. One famous example of this group is Taguchi designs \cite{Taguchi}, which utilize fractional facts of two, three, and mixed levels.

\subsection{Advantages of the framework}
\hfill \break

The main advantage of the factorial framework is to improve the sensitivity or velocity of experimentation compared to the traditional way of testing policies using $A/B/n$ testing, which makes it suitable for environments with time-varying effects. Furthermore, the framework allows us to make systematic assumptions about the nonexistence of interaction across factors.

\subsubsection{Improve sensitivity and velocity compared to policy A/B testing}
\hfill \break

In this section, we analyze the increase in the speed of experimentation that our framework provides. \\

Without breaking down the $n$ policies into factors, we typically assign all $I$ units equally under each policy, and one of the $n$ policies is used as the control or the baseline policy. In this setting, a natural way to estimate the treatment effect of any given policy is via a difference-in-means estimator, where we use the average of all units whose policy is the given policy minus the average of all units whose policy is the control policy. The variance of the estimator is $\frac{2\sigma^2}{I} n$, where $\sigma$ is the standard deviation of the outcome metric assumed to be the same across experimental groups in this discussion. \\

After breaking down policies into $F$ factors, we can estimate the treatment effect of any policy over the control policy as the sum of the difference between their associated factors. For the $f$th factor, the variance of the difference-in-means estimator between two different levels is $\frac{2\sigma^2}{I/L_f}$. Hence, the variance of the sum of these difference-in-means estimators is equal to $\frac{2 \sigma^2}{I}(\sum_{f=1}^{F} L_f)$. \\

From the above argument, we can get the ratio of the variances of the two estimators as follows:

\begin{equation} 
\frac{n}{\sum_{f=1}^{F} L_f} = \frac{\prod_{f=1}^{F} L_f}{\sum_{f=1}^{F} L_f} 
\end{equation}

Using the fact that the Minimum Detectable Effect (MDE) is linearly proportional to the standard error of the estimator, the ratio of the MDE of our framework to a typical policy A/B testing framework is $\sqrt{\frac{\sum_{f=1}^{F} L_f}{\prod_{f=1}^{F} L_f}} $. The ratio is less than $1$ when a factor has at least $2$ different levels, which means that our framework allows us to detect a much smaller effect than the policy A/B testing framework. In our use case where we break policies into $4$ factors with $2,3,2,2$ levels, respectively, we are able to detect an MDE which is $38\%$ of the MDE the the A/B testing framework.  \\

One can also look at the gain from the perspective of reduced sample size in order to detect the same MDE in our framework. A similar exercise like above shows that to get the same MDE, the ratio of the sample size required by our framework versus the A/B testing framework is $\frac{\sum_{f=1}^{F} L_f}{\prod_{f=1}^{F} L_f}$. In our empirical context, this implies a 267\% increase in the speed of measuring the same MDE.\\

Overall, the above discussion demonstrates that our framework can improve the experiment's sensitivity or velocity compared to a typical A/B testing framework. 

\subsubsection{Reduce experiment/policy setup cost}
\hfill \break

After we factorize the policy space, one way is to launch a full factorial experiment that tests all $\prod_{f=1}^{F} L_f $ combinations. However, practically this can involve high setup effort costs to select users targeting groups and implement them. In our framework, we reduce the number of policies needed to be actually tested by using fractional factorial design. This allows for a reduction in set-up costs, while still enabling estimation of the effects of all potential policies. 

\subsubsection{Suitable for a dynamic environment}
\hfill \break

Effects of policies can change over time due to time-varying environmental factors such as consumer types, inflation, employment rate, retail price, and other socioeconomic factors. Therefore, the optimal policy may change over time. Relative to the traditional sequential testing approach, our approach can help in such contexts by enabling quicker tests, discovering trends in effects, and allowing businesses to adjust their decisions accordingly.

\subsubsection{Comparison with a Multi-armed bandits approach}
\hfill \break

A typical multi-armed bandit framework  \cite{schwartz2017customer} improves experimentation efficiency by starting with an initial split of units across different experimental policies, learning about the performance of each policy as time goes on and updating the split of experimental units based on the performance. While we share the same goal of improving efficiency, our approach is different from the multi-arm bandit approach in the following ways.

\begin{itemize}
 
 \item A typical multi-armed bandits approach does not have an underlying model that could be used to extrapolate results from a subset of policies. Because of that, to learn about the treatment effect of the full policy space, the business needs to incur the full setup cost and experimentally launch all candidate policies. 

 \item Each iteration of multi-armed-bandits algorithm observes the metric value and then updates the traffic allocation accordingly. By running the iterations over time, the algorithm attempts to maximize the metric. This approach relies on the iterations completing in short time which enables a faster convergence to the optimal arm. This can be limiting in settings where the outcome of interest is longer term, as in our context where the outcome metric is 28-day conversion rate. Our approach does not have such a limitation because all experimental arms are evaluated once when the experiment ends.

 \item The multiarmed-bandits approach is more appropriate for problems in which one is mainly interested in optimization \emph{while} the experiment is running; it attempts to re-allocate the incoming sample at each iteration optimally based on what it has learnt up to the iteration. In contrast, our method is better suited to settings where the experiment is used to learn about a policy that is to be applied after the experiment ends \cite{10.1145/3534678.3539144}.
 
 \item \cite{scott2010modern} presents a fractional factorial bandit algorithm that takes advantage of fractional factorial deisgns. By assuming a parametric model, the parameter search space can be reduced relative to a typical MAB approach. However, similar to the most bandit algorithms, the goal is to minimize regret instead of accurately estimating and learning about the distribution of treatment effects of each variant, which is our goal.

\end{itemize}
\subsection{Estimate heterogeneous treatment effect}
\hfill \break

The above framework can be extended beyond estimating the average treatment effect for each policy to estimating the heterogeneous treatment effects of each factor level and, therefore, estimating the heterogeneous treatment effect of each candidate policy. When it is possible to conduct policy targeting based on the available experimental unit characteristics, this add-on benefit allows the business to assess and formulate personalized optimal policies. Let $X_i$ denote a vector of unit $i$'s pre-experiment characteristics. We can then extend our model to a more generic form. The model below uses a linear HTE model; as previously mentioned, a non-linear model can also be used.

\begin{equation}
    Y_i = \sum_{f=1}^{F}\sum_{l=1}^{L_f} W_{ifl}\left( \beta_{fl} +\lambda_{fl}'X_i \right) + \gamma'X_i + \epsilon_i \label{eq:HTEmodel}
\end{equation} 

Here, we introduce two new parameters compared to \eqref{eq:ourmodel}: $\gamma$ represents heterogeneity in the outcome $Y$ across units with different $X$s; and $\lambda_{fl}$ represents the corresponding heterogeneity in the treatment effect of policy factors.

\subsubsection{Validate if HTE model}
\hfill \break

If there are no strong priors of $X_i$s impacting the factor's treatment effects, one might statistically test the existence of systematic heterogeneity in the effects. \\

Specifically, we can test the null hypothesis that all interaction terms $\lambda$'s are equal to $0$;
\begin{equation}
H_0: \lambda_{11} = \lambda_{12} = ... = \lambda_{FL_F} = 0 \label{eq:hettest}
\end{equation}

If the null hypothesis is not rejected, there may not be detectable heterogeneous effects across the feature space. This may indicate that going from a $ATE$ to $HTE$ model may not lead to much improvement. However, this test can be conservative in practice \cite{feitBerman2019}, and firms might still use HTE if there are quantitatively significant benefits from it.

\subsubsection{Build HTE model}
\hfill \break

There are a variety of models available for estimating heterogeneous treatment effects. On a high level, they can be classified into direct and indirect estimation methods \cite{htemodel}.\\

Indirect estimation models are trained to minimize the loss function based on the observed and predicted metric values, such as the squared error. In our application that is: 

\begin{align}
L(\hat{\beta})=&E[(Y_i - \hat{Y_i}(X_i, \{ W_{ifl}, f=1,2...F, l=1,2,...L_f \}))^2] \notag \\
&=E\left[\left(Y_i - \sum_{f=1}^{F}\sum_{l=1}^{L_f} W_{ifl}\left( \beta_{fl} + \lambda_{fl}'X_i \right) - \gamma' X_i \right)^2\right]
\end{align}

Various approaches within the indirect approach are either parametric, putting different assumptions on sample distribution about the error terms $\epsilon$ and regularization (for example a linear or logit Lasso regression  model), or search for parameter values to minimize the loss function non-parametrically. \\

Having obtained an estimate of the regression above, we can predict the conditional average treatment effect based on the predicted expected outcome difference as:

\begin{align}
\hat{\tau}(X_i,  W^A, W^B) =  \sum_{f=1}^{F}\sum_{l=1}^{L_f} ( W_{fl}^A - W_{fl}^B) (\hat{\beta}_{fl} + \hat{\lambda_{fl}}'X_i )
\end{align}

where $W^A$ and $W^B$ are two sets of indicators representing two different policies $A$ and $B$ in the factor-level space; $\hat{\gamma}$, and $\hat{\lambda}$ are the estimated parameters; $\hat{\tau}(X_i,  W^A, W^B)$ denotes the estimated effect of switching from policy $A$ to $B$ for a unit $i$ with characteristics $X_i$. \\

The direct estimation models predict the conditional average treatment effect (CATE) directly. Say, we want to estimate the treatment effect of policy $B$ relative to $A$. We can use machine-learning methods to predict this effect. For example, we can use causal K-nearest-neighbors (KNN); for any user characteristics $X_i$, we find a set of $K$ nearest neighbors in the $X$ space which are given policy $A$ represented by $W^A$ in the factor-level space in the experiment, and similar set what was given the policy $B$ represented by $W^B$. We then estimate the CATE using the difference-in-means estimator:

\begin{align}
\hat{\tau}(X_i,  W^A, W^B) = \frac{1}{K} \sum_{u \in N_K(X_i, W^A)} Y_u - \frac{1}{K} \sum_{u \in N_K(X_i, W^B)} Y_u
\end{align}

Here $N_K(X_i, W^A)$ is the set of the K nearest neighbors with the policy $A$. Note that the estimator above is unbiased due to the independence between user characteristics $X_i$ and the policy assignment. Also, note that when the hyper-parameter $K$ uses the largest possible value,  $N_K(X_i, W^A)$ becomes all units received $A$, similarly, $N_K(X_i, W^B)$ becomes all units that received $B$. The estimator becomes equivalent to the estimator given by (7). \\

To tune hyper-parameter $K$, a loss function based on the actual treatment effect is infeasible because we do not observe the true treatment effect. Hence, we can use transformed outcome loss. We first transform the metric $Y_i$ to $Y_i^*$ as

\begin{equation}
Y_i^* = \frac{W_i-Prob(w = W^A |  X_i)}{Prob(w = W^A |  X_i)(1-Prob(w = W^A |  X_i))}Y_i
\end{equation}

Given that the transformed outcome is an unbiased estimator of CATE given unconfoundedness \cite{hitsch2018heterogeneous}, we can choose $K$ by minimizing the squared error between $Y_i^*$ and $\hat{\tau}(X_i,  W^A, W^B) $

\subsection{Evaluate the framework}
\hfill \break

There are two parts of evaluations that we conduct, one to evaluate the correctness of our framework to test the ATE of each policy, which essentially evaluate the correctness of our additive main effects and no-interaction assumptions; the other one to evaluate the HTE framework.  In the evaluation of the HTE framework, we not only want to evaluate the prediction accuracy but more importantly, we want to evaluate how much the key business metrics such as profit can be improved using the framework. \\

\subsubsection{Evaluating Model Assumptions with Out-of-Sample Variant } \label{subsubsec:eval}
\hfill \break

To evaluate the correctness of our framework, we launch an additional variant formed by the same set of factors but not belonging to the other experimental test policies (we refer to this as the out-of-sample or held-out variant). From the experiment, we can estimate the treatment effect of the out-of-sample variant relative to each in-sample variant via a t-test. From there, we can jointly test whether the observed treatment effects and our predicted treatment effects from our framework are statistically different.

\subsubsection{Evaluating Model Assumptions with Out-of-Sample Variant on User Segments}
\hfill \break

To get more data points for comparison, especially since the noise-signal ratio is high, we propose the following additional comparisons. First, we can compare the out-of-sample variant with each of the in-sample-variant and compare the predicted differences with actual observed differences. \\

Additionally, we repeat this procedure across various distinct consumer groups. The idea is to split users into heterogeneous groups. Once the split is done, we can compare the predicted treatment effects with the observed ones for each group. If our framework holds, we expect the observed and predicted effects to match across all groups.

Here is an overview of the procedure:

\begin{enumerate}
\item Select user features $X_i$ which are historically known to predict the effects of similar treatments, or those which are highly predictive of metric $Y_i$

\item Train a regression or another machine learning model $f: X_i \rightarrow Y_i$ using historical data and apply $f$ to predict on the experimental sample as well as the sample corresponding to the out-of-sample variant.

\item Choose $K$ as the number of groups, and discretize the sample into equal-sized $K$ groups based value of the predicted outcome. The number of groups $K$ can be chosen by any cluster analysis algorithm such as partitioning clustering like K-means or density-based clustering like DBSCAN.

\item For each sample subgroup and each experimental variant, we calculate the difference between the average predicted outcome value for those sample units assigned the experimental variant and those assigned the out-of-sample variant. This is the predicted relative treatment effect. 

\item  Similarly, for each sample subgroup and each experimental variant, we calculate the difference between the average observed outcome value for those sample units assigned the experimental variant and those assigned the out-of-sample variant. This is the observed relative treatment effect.

\item Compare the predicted relative treatment effects 
with the "true" observed relative treatment effect across all the groups and variants.

\end{enumerate}

\subsubsection{Evaluation based on Optimal Prediction}
\hfill \break

For each instance in the characteristics space, $X_i = x$, our framework can predict the outcome from adopting different policies represented by $\{ W_{ifl}, f=1,2...F, l=1,2,...L_f \}$. Hence across all predictions we are able to identify the optimal policy that maximizes the desired outcome. Denote the opitmal policy by $ W^{*}_i $,\\

\begin{equation}
  W^{*}_i = \arg \max_{W_{ifl}}  \hat{Y_i}(X_i, \{ W_{ifl}, f=1,2...F, l=1,2,...L_f \})
\end{equation}

The challenge of the evaluation is that we do not observe unit $i$'s actual outcome under policy $W^*_i$. We observe the ``ground truth'' data for some, but not all units. \\

In this and the next two subsections, we suggest alternative ways to assess the value of using this framework, under different assumptions. \\

One simple way is as follows. For each feature, after we select the optimal policy $W^{*}_{i}$ that maximizes the predicted metric value, we can also record the prediction associated with the optimal policy $\hat{Y_i}(X_i,  W^{*}_i)$. With those, we can simply evaluate the model performance by taking the average of the metric across all users.

\begin{equation}
\sum_{i \in D_{test}} \hat{Y_i}(X_i,  W^{*}_i) / |D_{test}|
\end{equation}

$D_{test}$ is the set of experimental units in the data. The advantage of this evaluation is that it utilizes all the data we have. However, because the evaluation is purely based on predictions, it requires the model to be unbiased to get a reasonable evaluation.

\subsubsection{Evaluation based on adjusted Expected Response Under Proposed Treatments (ERUPT)}
\hfill \break

The second evaluation procedure only considers users whose assigned policy happens to be the same as the predicted optimal policy by chance. The evaluation is given by

\begin{equation}
\sum_{i \in D_{test}} \frac{Y_i I\{ W_{i} = W^*_{i}\}}{e(X_i,W^*_i)} / |D_{test}|
\end{equation}

where $e(X_i,W^*_i) = \Pr(W_{i} = W^*_{i}  | X_i)$. \\

As shown from above, it takes the observed metric value of each user whose observed policy is the same as the model suggested policy and inversely weights the metric with the targeting probability based on feature $X_i$. When the observational data comes from a uniformly randomized experiment, the targeting probability becomes a constant $e(X_i) = e$. In this case the adjusted ERUPT gives the same evaluation as the simple average across users whose observed policy is the same as the model suggested policy. \\

However, this procedure has the following limitations:

\begin{enumerate}

\item  We only have a limited number of policies in an observed fractional factorial experiment, while our proposed policies can come from the entire policy space. Hence this evaluation may be inaccurate because it is based on the recommendations of a subset of the data.

\item Even when we have all policies in the observed policies set, the number of users whose observed policy is the same as the model suggested policy can be quite small, which can make the results highly imprecise.

\end{enumerate}

\subsubsection{Evaluation using a new randomized experiment}
\hfill \break
The most reliable way to evaluate the impact of this procedure is to launch an additional randomized experiment, which can be costly compared to the above approaches. On a high-level, this test randomly splits users into two groups. With one group of users adopting the optimal personalized policy suggested by the HTE model; and another group of users adopting the baseline policy. \\

\subsection{Launch optimal variants in framework} \label{sec:launchoptimalvariant}
\hfill \break

After analyzing the experiment data, we need to decide what to do next in order to optimize the metric of interest. Based on prior learning, quantitative benefits from personalization, and statistical analysis such as results from the interaction test result (\ref{eq:hettest}) one can decide whether a personalization would add incremental business value. In practice, launching the personalized model might be viable only when there is a significant heterogeneity in effects, given the much higher operating and engineering cost to maintain the personalized model.
\\

If maintaining a personalized model does not provide a net benefit we can apply the model (\ref{eq:ourmodel}), get an estimate for each $\hat{\beta}_{fl}$, and define the single optimal policy which sets the level of each factor as

\begin{equation}
f_{l^{*}} = \arg \max_{l \in (1,2,...L_f)}  \hat{\beta}_{fl} \quad (f = 1,2,...F)
\end{equation}

Alternatively, we can apply the personalization model (\ref{eq:HTEmodel}) and get an estimate for each  $\hat{\beta}_{fl}$ and  $\hat{\lambda}_{fl}$, then we can define the optimal policy for each user characteristics $X_i$ as

\begin{equation}
f_{l^{*}} | X_i = \arg \max_{l \in (1,2,...L_f)}  \hat{\beta}_{fl} + X_i \hat{\lambda}_{fl} \quad (f = 1,2,...F)
\end{equation}

\subsection{Summary of our Framework and Proposed Approach}
\hfill \break

Our framework breaks down the policy space into factors, and each factor can take one of many levels. This allows us to cast our policies into a vector space and use factorial experiment designs to estimate the effect of different factor levels, and then the effect of different potential policies. The main benefit of our framework is to improve the velocity of learning, which can be measured by the improved minimum detectable effects (MDE) of policies given a fixed sample, or by the smaller sample size required to test a certain number of policies. Applying fractional factorial designs can simplify the experiment design and save set up costs which are prohibitive in typical business contexts. \\

In many contexts we expect heterogeneous effects of policies across experimental units. Our framework can be extended to account for such heterogeneity and can be augmented with machine learning techniques to estimate personalized policy recommendations. \\

Our approach suggests tests to verify the validity of the framework, particularly the additivity assumptions one might make in simplifying the experiment, and evaluating its overall impact. \\

\section{Business Context} \label{sec:businessCont}
\hfill \break

We focus on an application of our approach at DoorDash's Consumer Retention Marketing. Doordash's primary offering ``DoorDash Marketplace'' provides a suite of services that enable merchants to establish an online presence, generate demand, seamlessly transact with consumers, and fulfill orders primarily through independent contractors who use its platform to deliver orders.\footnote{DoorDash Inc. (2023) Form 10-Q. U.S. Securities and Exchange Commission. https://d18rn0p25nwr6d.cloudfront.net/CIK-0001792789/83885643-4f88-4a45-a869-cd2fef24524e.pdf} The DoorDash Consumer Retention Marketing team aims to build a lasting relationship with customers as soon as they engage with DoorDash by providing them useful personalized experiences and driving them to return to DoorDash to find the relevant merchants and products. \\

Targeted promotional campaigns are commonly used to improve consumer retention and are used at DoorDash as well. Promotional offers may not be one-size-fit-all, for example, consumers who always order over the weekend may find a promotional email coming in on a Friday night more timely and actionable; consumers who order smaller basket size may find promotion without minimal avg-order-spend requirement more favorable. From a business standpoint, the available marketing budget for such promotional campaigns is limited so assessing and improving promotional campaigns is important, and usually done using randomized experiments. \\

\subsection{Status Quo Experimentation Approach and Limitations} \label{sec:statusquoapproach}
\hfill \break

Incremental value of a marketing policy, in general, is estimated using a randomized field experiment whenever possible at DoorDash. When optimizing a marketing program, analysts iteratively experiment on hypotheses, ship winning variant, conduct dimension analysis based on the experiment's data -- analyze how different segments within the sample respond differently, based on which analysts hypothesize further improvements, and propose the next round of experiments. \\

There are some key challenges that significantly limit the experimentation and innovation speed: 
\begin{itemize}
\item Experiment duration for running sequential testing can be long due to factors such as limited testable population, limited marketing budget which limits the total sample for testing, and long term success metrics which require long measurement period. 

\item If we were to conduct a randomized experiment testing all possible combinations in one experiment, both the direct implementation cost (i.e. labor cost and marketing spend) and indirect opportunity cost of exposing consumers with sub-optimal policies could be extremely high \cite{htemodel}. On labor cost, implementing a multi-variant marketing experiment takes an average 1 week per variant to complete the following key steps: 
\begin{itemize}

    \item Determining target audience (i.e. who is eligible to get this campaign)
    \item Configure the promo code (i.e. what is the promo code, minimum avg-order-spend requirement, \%off or \$off, incentive amount, maximum redemption count, eligible market and so on)
    \item Design content copy for off-app channels (e.g. email and push) and in-app placements (e.g. home page banners, carousels)
    \item Build an automated workflow to send off-app notifications (i.e. when to trigger announcement and reminder notifications)
    
\end{itemize}

This roughly means 7-8 operator week for setting up an experiment with 8 variants, 24 operator week for setting up an experiment with 24 variants. We believe this estimate is conservative given more QA time will be needed because the implementation is more prone to human error when number of variants is higher. Marketing spend is also a key consideration when running monetary marketing experiments. It is fair to assume marketing cost will linearly scale with the number of variants we test holding statistical power constant. The indirect cost refers to the opportunity cost of offering less performing treatment during the entire experiment duration. 

\item There are time-based confounders when running sequential testing due to external factors and seasonality. This makes comparing performance of treatment tested at different time difficult. In fact, same marketing offers exhibiting different levels of engagement and cost-effectiveness occurs quite often based on marketing experiments conducted at DoorDash.

\item Running longer experiments also exposes the work steam with potential risks of not being able to ship the optimal feature and hinder future experiments (campaign in our case) should there be business focus shift resulting in situations such as a budget pull back.

\end{itemize}

\section{Applying our Framework}
\hfill \break

With the approach proposed in this paper, we can practically and rigorously speed up the business decision. The next section describes our approach in detail. Overall, we take the following steps:

\begin{itemize}
\item Factorization. We view a promotion campaign has the following four factors: Promo spread, Depth of discount, Triggering time, and Messaging. Each of the four factors has two or more levels. For example, Promo spread can be “Upfront” (i.e. inform consumers about the full length of the promo upfront)  or “Spread” (i.e. break a multi-week long promo into several weeks and inform consumers sequentially)

\item Fractional Factorial Experiment Design. Based on the total number of feature combinations, we generate an experiment design that is efficient in the sense that it can predict the impact of any feature combination while actually testing a few combinations. In our case there are total 2 $\times$ 3 $\times$ 2 $\times$ 2 = 24 variants whose effects we want to estimate. By running the fractional factorial design, we are able to reduce the total number of the variants to 8, that we actually test at scale. 

\item Include an additional out-of-sample variant in the experiment to test the ``no-interaction'' assumption we make in our model. We compare the observed effect of this variant with the model-predicted treatment effect to assess the validity of the model assumptions.

\item Launch the experiment based on the variants generated in the previous step and gather data up to a sample size which is pre-computed based on the minimum detectable effect.

\item  Estimate the average effect of each feature: Analyze using all features as regressors and our main outcome metric as the independent variable. From the estimated coefficient for each factor-level, we are able to derive the treatment effect for all possible promo policy permutations.

\item HTE: Use machine learning to estimate heterogeneous treatment effects that enable us to predict, separately, the impact of each potential feature combination on each customer profile. Based on these predictions we can personalize promotions to each customer.

\end{itemize}

\section{Experiment design } \label{sec:expDesign}
\hfill \break

The targeted promotional campaign to which we applied this framework is an evergreen multiweek-long promotion that has gone through many iterations. On the basis of the previous experiments, the team developed hypotheses for the next set of changes to improve the results. For example, redemption often occurs when the promotion is first announced and when it is about to end. The team hypothesized that gradual release of the promotional benefit instead of ``unlimited free delivery for x weeks straight'' could drive a sense of urgency at any point in time more evenly and drive sustainable impact on consumer behavior. Another example is that there are consumers who typically place orders on weekdays versus weekends. Therefore, communicating these promotions on the weekend versus on weekdays could lead to materially different results for the same type of audience. \\

These hypotheses, which would have been tested sequentially and suffered from the limitations listed in Section \ref{sec:statusquoapproach}, inspired us to construct the factors discussed in the next section. 

\subsection{Break down campaign design into factors}
\hfill \break

\begin{figure}[h]
\centering
\includegraphics[width=0.4\textwidth]{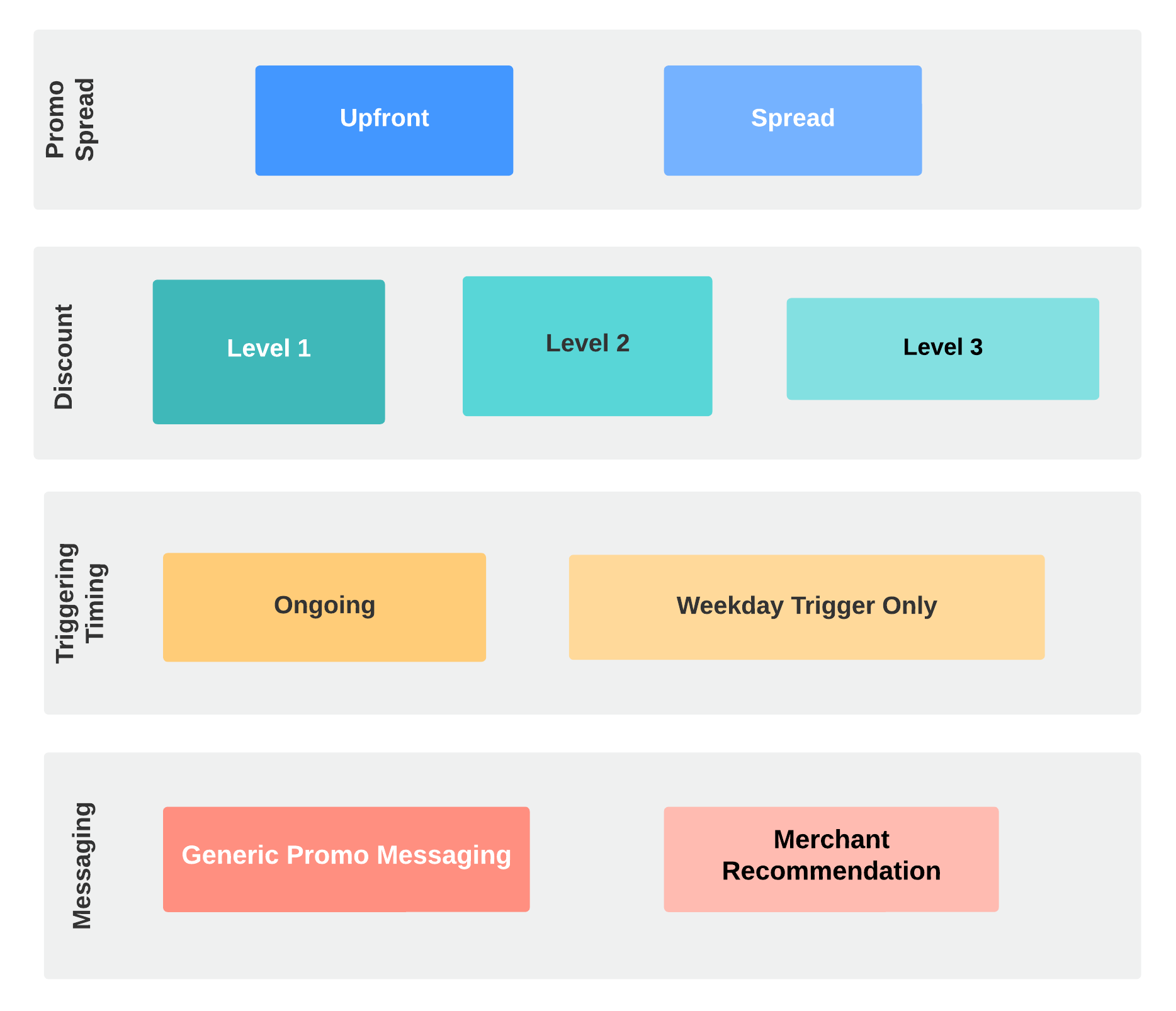}
\caption{Selected factors} \label{fig:components}
\end{figure}

In order to design our treatments in a more methodical way, we identified four main factors that are at the core of our hypotheses: promo spread, discount, triggering timing and message. Figure \ref{fig:components} shows our factors and levels. Below we describe them in detail.\\

\begin{itemize} 
\item Promo Spread: There are many ways one can deliver a long-duration, say a 4 week, promotion. A straightforward and commonly used way is to provide and communicate the whole promotion incentive upfront to consumers via email or mobile push notifications. The consumers can then use their promotion any time in the next 4 weeks. We call this "Upfront". This could drive attention and lead to a spike in ordering when we first announce it. However, that initial hype may not last long and may not drive sustainable engagement. As opposed to Upfront, Spread promo is where the same offered incentive is divided into two halves that are given in sequence. So consumers receive two 2-week long offers; one in week 1, and another in week 3. This can change the amount of attention consumers give to the incentive and times when they do so. 

\item Discount: Discount is the type of monetary incentive we can offer. It can take various formats such as ``free delivery'', ``percentages off'', ``dollars off'' or a combination of them. In this experiment, we picked three discounts with similar expected value but different formats towards which consumers could have different perceptions.  The total discount value was decided based on historical experiments on the same targeted population. 

\item Triggering Timing: Triggering timing is \emph{when} we trigger the marketing campaign to consumers. Campaigns may be triggered by consumer events, for example, when consumers experience a bad delivery, or consumers stop ordering for an extended period of time. There are two types of triggering timing of interest here: 
\begin{itemize} 
        \item Ongoing: Start the campaign as soon as the consumer becomes eligible
        \item Weekday trigger only: trigger only on weekdays. For consumers who happen to become eligible over the weekends, we trigger the campaign on the next Monday. 
     \end{itemize}

\item Messaging: Email and push notifications are the primary channels used to communicate offers to consumers. While push notification usually contains short messages, email content can be rich, which gives two options: 
\begin{itemize} 
        \item Generic promo messaging: the email speaks only about the promotion being offered, which is a relatively plain email.
        \item Merchant Recommendation: the email provides some recommendations to the consumer in addition to the offer. 
     \end{itemize}

\end{itemize}

\subsection{Apply Fractional Factorial Design}
\hfill \break

After creating these four factors, three of which have two levels and one has three levels, we have 24 combinations. So a full factorial design testing all combinations will require 24 experimental arms. There are major practical operational challenges in setting up such a 24-arm marketing campaign, as mentioned in section \ref{sec:statusquoapproach}.  \\

\begin{figure}[h]
\centering
\includegraphics[width=0.4\textwidth]{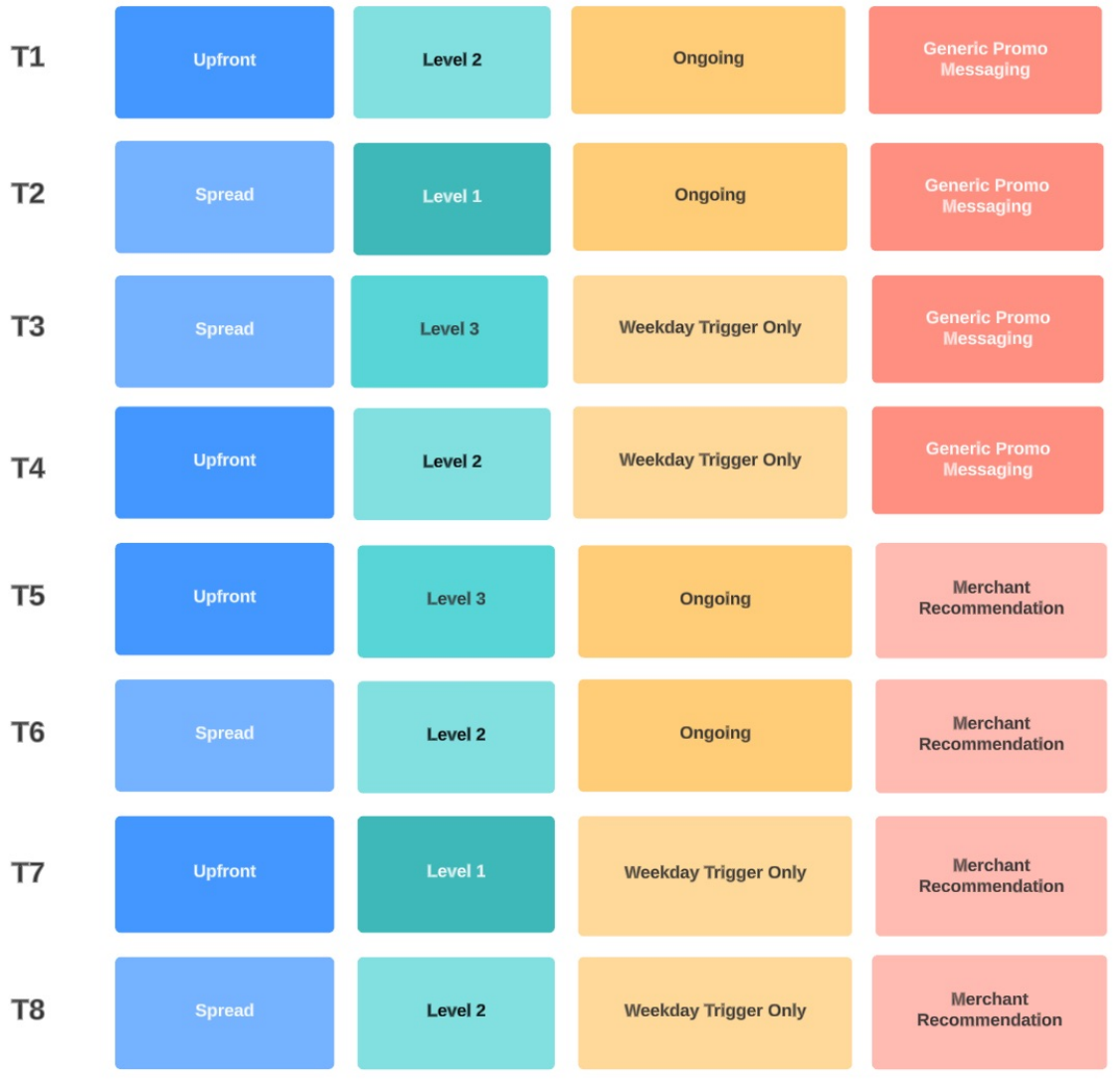}
\caption{Experimental Treatment Arms Implemented} \label{fig:DOE}
\end{figure}

To solve this problem, we apply fractional factorial design shown in figure \ref{fig:DOE} to shrink the number of variants from 24 to 8, which makes the execution manageable while retaining the ability to make inferences about the untested variants by making modeling assumptions. In addition, when factors are independently randomized, we are able to effectively use the whole sample to analyze the impact of each factor.\footnote{Additionally, we include in our design a randomly chosen Control group of users who are not given any promo. The sole purpose of the Control group is to provide a benchmark to estimate the overall impact of the program.} \\

\subsection{Additional Experimental Arm for Validation}
\hfill \break

Evaluation of the 16 policies excluded from the experiment relies on modeling assumptions of no significant interaction terms. Such effects are likely to be small, according to the team's priors but no theory guarantees this. To empirically test this, we launch a 9th experimental arm, referred to as the ``out of sample'' arm, which is randomly chosen from the excluded 16. The idea is to compare the observed effect of this variant relative to others, and the corresponding model predicted effect to validate our framework's assumptions. The 9th arm was implemented after the other eight, and ran concurrently for a subset of the time period. So for validation comparisons below, we will use a subset of the data when all arms were concurrently active. \\

\subsection{Data}
\hfill \break

\subsubsection {Metric selection}
\hfill \break

The marketing campaign's goal is to encourage more usage of the product hence driving higher sales volume while spending promotion budget efficiently. Therefore, our main success metric is ``average profits per user'' which is sales revenue minus costs, such as promotional costs within 28 days after first exposure to the promo.  For the purpose of validating our framework we use an additional metric, order volume or order rate, for robustness. While these are the two metrics we use for analysis, the company tracks other metrics such as customer retention and promo redemption rate, and guardrail metrics such as delivery quality, and manual support ticket volume (which would surge if a promotion is not implemented properly). \\ 

Viable Policy. The main success metrics are used to make projection about the payback period: how long it will take for the promotion spend to be paid back in full. A marketing policy is considered to be viable for scaling if its payback period is within an acceptable range, provided no statistically nor economically significant change in guardrail metrics. For the results section of the paper, we have used this metric. \\

\subsubsection {Sample Size Calculation}
\hfill \break

As in a traditional A/B test, we assess the required sample size by calculating the baseline of our outcome metric, estimating the minimum detectable effect (MDE) and calculating the smallest sample size required at the chosen type I (5\%) and type II error rate (20\%). \\

Our experimental design enables us to test the comparative effectiveness of multiple factors, each of which has two or more variants \cite{powerful}. In other words, we are testing different levels within each factor against the base level of that factor. For example, how does a Spread promo perform compared to Upfront. Therefore, when estimating the MDE, we want to estimate the MDE for each of the four factors, as we expect different sensitivity for different factors based on our historical experiments and domain knowledge. \\

Using the Promo Spread factor as an example, we first use business tradeoffs to choose a minimum detectable effect (\textit{d}) of the success metric and calculate the sample size needed for each level of this factor using the following formula:

\begin{equation}
n = \left[ \frac{(z_{1-\alpha}+z_{1-\beta})\sigma}{d} \right]^2
\end{equation}

where $\sigma$ is the standard deviation of the dependent variable. Given that Promo Spread has two levels, the total sample size needed is \textit{n} $\times$ 2. We repeat this sample-size calculation process for the other three factors and go with the largest sample size required.

\subsubsection {Covariates}
\hfill \break
Past experiments and customer data analysis show customer heterogeneity in response to promotional offers. In the team's experience, a consumer's previous average behavior is also predictive of his future behavior. On the basis of this understanding, we chose covariates to assess heterogeneity in effecs. Here are some examples of chosen covariates:
     \begin{itemize} 
        \item Lifetime orders. The amount of orders a consumer previously placed on DoorDash typically serves as a signal of product adoption and loyalty. Consumers who have placed orders beyond a certain threshold are more likely than others to adopt new product features, and react to marketing messages and redeem promotions.
    
        \item Promotion usage. Historical promotion usage is an indicator of customer price sensitivity and the likelihood of promo redemption when receiving one. 
        
        \item App visit pattern. Frequent app visit without placing orders could indicate promotion seeking or price comparison.
        
        \item Basket size. Historical basket size (average spend per oder) is observed to correlate with profitability. Customers who tend to place bigger sized orders could redeem more promotions especially when given "\%off" type of promotions. This is an important feature to consider when analyzing a promotion's benefit. 

        \item Order day of the week. Whether the order arrives during weekdays or weekends indicates the role DoorDash plays in the customer's life style. Weekday orderers may view food delivery service as a way to bring convenience into their fast paced lifestyle while customers who tend to place orders over the weekend may crave for something but don't want to leave their home and wait in line. Such customer types may respond differently to promo configurations.

     \end{itemize}

\section{Results}
\hfill \break

In this section, we discuss the results of our analysis of the experiment. We first test the validity of our framework by comparing our predictions of the out-of-sample variant with the observed outcomes. Then we assess the heterogeneity in treatment effects. Lastly, we show the business impact of our approach: we determine the optimal policy, the importance of the factors considered, and the role of heterogeneity. To preserve confidentiality, we have multiplied all the metric values and treatment effects with an undisclosed constant.

\subsection{Testing Model Assumptions}
\hfill \break

\subsubsection{Joint Test on Out-of-Sample Variant}
\hfill \break

As we described in Section \ref{subsubsec:eval}, we conducted a joint test to check if there is a statistically significant difference between our framework's predictions and observed data. Given that we have a total of eight in-sample variants and one out-of-sample variant, there are eight differences in total; $\forall i$ predicted profit change going from arm $i$ to out-of-sample arm, denoted as arm 9, versus the observed change. As a result of the joint test of differences, we obtain a p-value of $0.80$, which means that there is no detectable difference between the predictions of our framework and the observed data. This supports the validity of our framework end-to-end, especially our assumption of no interaction effects. \\

\begin{figure}[h]
\centering
\includegraphics[width=0.4\textwidth]{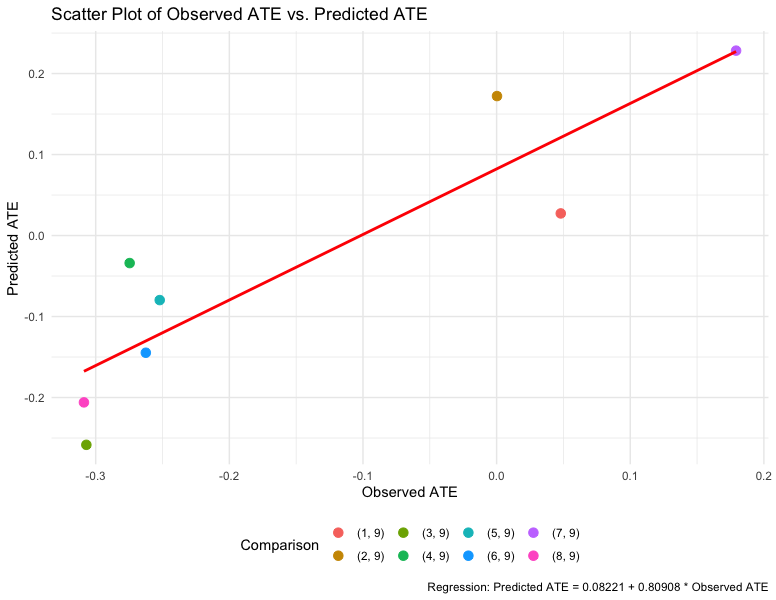}
\caption{Design Validation - Metric: VP} \label{fig:Valid_vp}
\end{figure}

\begin{figure}[h]
\centering
\includegraphics[width=0.4\textwidth]{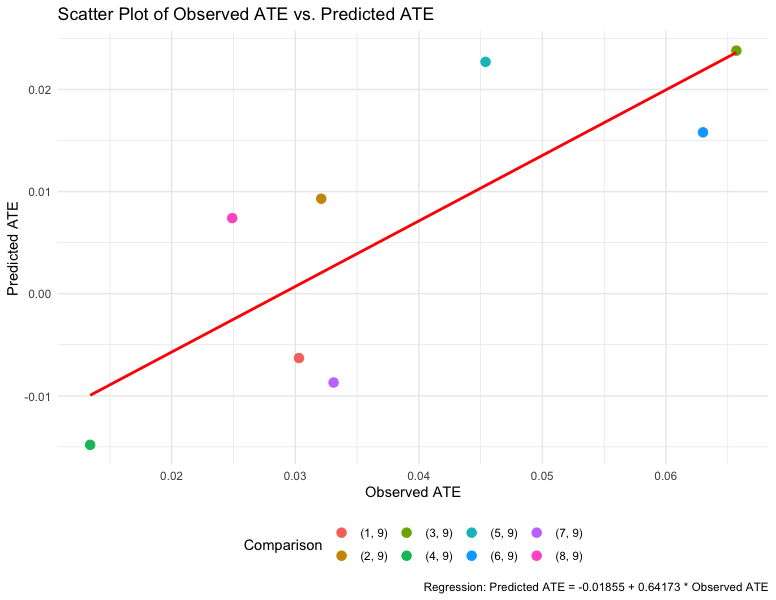}
\caption{Design Validation - Metric: Order Rate 14d} \label{fig:Valid_order}
\end{figure}

Figure \ref{fig:Valid_vp} plots the predicted and observed effects for a visual comparison, showing that the predicted and observed effects are significantly correlated. For robustness, we repeat this analysis using orders as our dependent metric. Figure \ref{fig:Valid_order} shows a similar high degree of correlation between the actual observed and predicted effects. Overall, this analysis supports our assumption that the model is capable of predicting results in policy configurations not included in the experiment.
\\

\subsubsection{Evaluate across User Segments}
\hfill \break

To conduct further comparisons using heterogeneous user populations, we grouped the sample by features such as average spend per order (avg-order-spend), promotion lifetime orders, orders rate during pre-churn period, churn tenure, and number of visits during the pre-churn period. In total, we chose the number of groups $K = 10$, so in total we have $10 \times 8 = 80$ predicted effects to compare with the observed effects. \\

\begin{figure}[h]
\centering
\includegraphics[width=0.4\textwidth]{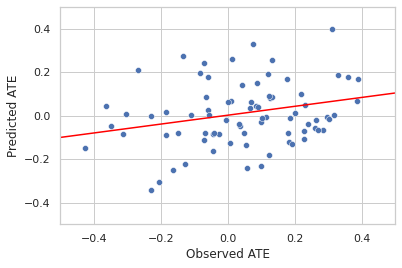}
\caption{Compare predicted HTE to observed HTE \\ - Metric: Profit} \label{fig:HTEValid_vp}
\end{figure}

\begin{figure}[h]
\centering
\includegraphics[width=0.4\textwidth]{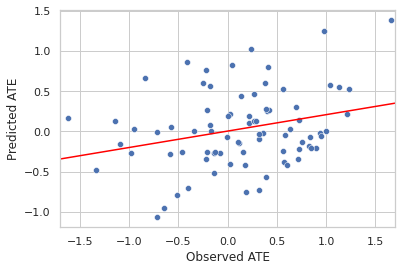}
\caption{Compare predicted HTE to observed HTE \\ - Metric: Orders} \label{fig:HTEValid_orders}
\end{figure}

From the plots in figures \ref{fig:HTEValid_vp} and \ref{fig:HTEValid_orders} we can visually see that the predicted and the observed user-segment level treatment effects are positively correlated. The slopes in both these figures are statistically significant (corresponding $p$-values equal to 0.02 and $<$0.01 respectively) which further supports our framework.\\

It is also interesting to note that the variance of the predicted treatment effects is less than the variance of the observed ones, which occurs due to the additional sampling noise in the observed data.
\\

This suggests that this evaluation method may have a higher false-positive rate when the unexplained variance is high.\footnote{In other contexts, researchers have proposed adjustments of the prediction variance by accounting for the prediction error \cite{Duan_2021}.}

\subsection{Factorial Experiment Results}
\hfill \break
We begin our analysis by estimating the impact of our factor levels on the business outcome metric, average profits per user, by estimating the $\hat{\beta_{fl}}$'s that are the estimates of the parameters in \eqref{eq:ourmodel} using a linear regression. Table \ref{reg:factormodel} shows these estimates, with the first level of each factor as the omitted baseline. \\

We assess the importance of each factor, which is the maximum impact the factor can have on the outcome metric. Specifically, for a factor $f$, we calculate $\max_l\{\hat{\beta_{fl}}-\min_l\hat{\beta_{fl}}\}$. By this measure, we note that Discount is the most important factor, followed by Promo Spread, Messaging, and Trigger Timing. Within the Discount factor, Level 3, which stands for a ``\%off'' with a limited redemption count promo  representation of the discount has the largest and most statistically significant coefficient. \\

Overall, these estimates tell us that the way the discount is communicated in a promotion is the most important factor among those considered here, and, specifically, a unified ``\%off'' with a limited max redemption count representation of the discount has the greatest impact on profit for the target audience of the program we optimize for. \\

Next, we use the approach described in Section \ref{sec:launchoptimalvariant} to predict a policy based on the combination of the best level $f_{l^*}$ of each factor $f$. Based on this calculation, our optimal policy is: Upfront Promo Spread; Discount conveyed as a ``\%off'' with a limited max redemption count; Ongoing Triggering; Generic Messaging. This policy happens to be out of sample, that is, it was not included in the eight-arm experiment. Its predicted profit is greater than the highest among the eight experimental arms by 1\% and higher than the control group by 5\%. Table \ref{tab:policypred} in the Appendix shows our predicted profits from each of the 24 possible policies. \\

\begin{table}
\tiny
\centering
\begin{tabular}{m{3.5cm}cccc}
           \textbf{Variable Name}                                    & \textbf{Coef} & \textbf{Std err} & \textbf{t} & \textbf{P$> |$t$|$}\\
\midrule
Intercept                              & 7.35231 & 23.08625 &     61.032   &          0.000        \\

Promo Spread [Upfront]            & 0.12905 & 0.40522 &      1.414   &          0.157             \\

Discount [Level2]                  & 0.40757 & 1.27977 &      3.648   &          0.000            \\

Discount [Level3]                  & 0.41950 & 1.31723 &      3.253   &          0.001        \\

Trigger Timing [weekday]  & -0.01350 & -0.04239 &     -0.147   &          0.883        \\

Messaging [Merchant Recs]                  & -0.02732 & -0.08578 &     -0.298   &          0.766\\
\bottomrule
 \multicolumn{5}{l}{Notes: Variables are represented as ``Factor[Level]''.} \\
 \multicolumn{5}{l}{We use data on the eight experimental arms for this estimation.}
\end{tabular} 
\caption{Regression of Average profits per user on factor levels.}
\label{reg:factormodel}
\end{table}

\subsection{Heterogeneous Treatment Effects}
\hfill \break

When we conduct a joint test \eqref{eq:hettest} of interaction between the factors and all user characteristics, we are unable to reject the null hypothesis ($p$-value $=0.2$). More detailed results are attached in the appendix, where we can see most interaction terms are not statistically significant. This analysis indicates that a blanket approach of detecting heterogeneity may not be suitable for our application.\\

Based on findings from previous campaigns, there might still exist some user-level heterogeneity that can impact business outcomes. To investigate this possibility more closely, we pick one feature that has the highest historical correlation with our outcome metric; avg-order-spend, which is the average amount of money in dollars a user spent on previous orders. We regress our outcome metric on factor levels, interacting them with avg-order-spend.  Table \ref{reg:interactionmodel} shows the results from this regression. Notice that several interaction terms, such as those with trigger timing and discount, are statistically significant (joint test $p$-value = $0.01$). \\

This heterogeneity recommends different optimal policies across users with different avg-order-spend. For example, controlling for Discount, Promo Spreat, and Messaging, the treatment effect of Trigger Timing [Weekday] relative to Trigger Timing [Ongoing] is $0.3636 - 0.0147 \times$ avg-order-spend, which means when avg-order-spend is less than about \$25, Trigger Timing [Weekday] is more profitable; the opposite is true when avg-order-spend is greater than \$25. This differs from the recommendation of launching a blanket ongoing trigger timing made by the model without heterogeneity. \\

We  present the optimal policy result in Table \ref{tab:htepolicy}, where we discretize the avg-order-spend as 0, 1, 2, etc., recommend different policies, and give different predicted profits given different ranges of the avg-order-spend. From the results, we can also see that the optimal arm selected in Table \ref{tab:policypred} is only  optimal in Table \ref{tab:htepolicy} when the avg-order-spend is between 25 and 26.\\

Using our dataset to compare the predicted benefits from using HTE we find that the HTE model can generate $2\%$ more profit.

\begin{table}
\tiny
\centering
\begin{tabular}{m{3.5cm}lcccccc}
\textbf{Variable Name} & \textbf{coef} & \textbf{std err} & \textbf{t} & \textbf{P$> |$t$|$} \\
\midrule
Intercept                                                          & 1.25977 & 0.21980 &      5.734   &          0.000  \\

Promo [upfront]                                         & 0.26784 & 0.16642 &      1.602   &          0.109  \\

Discount[level3]             & 0.93321 & 0.23550 &      3.951   &          0.000   \\

Discount[level2]           & 0.63302 & 0.20410 &      3.089   &          0.002     \\

Trigger Timing[weekday only]                            & 0.36361 & 0.16642 &      2.176   &          0.030      \\

Messaging[mxrec]                                           & 0.01256 & 0.16642 &      0.076   &          0.940    \\

avg-order-spend                                                   & 0.23864 & 0.00628 &     33.103   &          0.000      \\

Promo[upfront]$\times$avg-order-spend                          & -0.00534 & 0.00628 &     -0.998   &          0.318  \\

avg-order-spend$\times$Discount[level3]  & -0.01947 & 0.00628 &     -2.520   &          0.012      \\

avg-order-spend$\times$Discount[level2]  & -0.00848 & 0.00628 &     -1.237   &          0.216       \\

Trigger Timing[weekday only]$\times$avg-order-spend                 & -0.01476 & 0.00628 &     -2.681   &          0.007       \\

Messaging[mxrec]$\times$avg-order-spend                               & -0.00126 & 0.00628 &     -0.212   &          0.832      \\
\bottomrule
 \multicolumn{5}{l}{Notes: Variables are represented as ``Factor[Level]''.} \\
 \multicolumn{5}{l}{We use data on the eight experimental arms for this estimation.}
\end{tabular} 
\caption{Regression of Average profits per user on factor levels interacting with avg-order-spend.}
\label{reg:interactionmodel}
\end{table}

\begin{table}
\centering
\resizebox{\columnwidth}{1.1cm}{
\begin{tabular}{lllllr}
\toprule
avg-order-spend & \multicolumn{4}{c}{Factors} & Predicted Profit  \\
& Promo Spread & Discount & Trigger Timing & Messaging & \\
\midrule
$[0, 10]$ &      upfront &   level3 &      weekday only &   mxrex & $2.837 + 0.1975 \times avg-order-spend$ \\
$[11, 24]$  &      upfront &   level3 &      weekday only &   generic & $2.824 + 0.1986 \times avg-order-spend$\\
$[25, 26]$ &      upfront &   level3 &           ongoing &   generic & $2.461 + 0.2134 \times avg-order-spend$\\
$[27, 48]$   &      upfront &   level2 &           ongoing &   generic & $2.160 + 0.2247 \times avg-order-spend$\\
$[49, 75]$  &       spread &   level2 &           ongoing &   generic & $1.893 + 0.2301 \times avg-order-spend$ \\
$[76, \infty]$  &       spread &   level1 &           ongoing &   generic & $1.260+ 0.2385 \times avg-order-spend$\\
\bottomrule
\end{tabular}
}
\caption{Optimal Policies given avg-order-spend}
\label{tab:htepolicy}
\end{table}

\section{Conclusion}
\hfill \break

Businesses have begun to rigorously use A/B testing to optimize policies. However, the path to optimization using A/B testing requires effort, is time-consuming, and necessitates prioritization over potential hypotheses to test in typical settings with low statistical power.  This paper presents a case study with empirical experiment data that conducts and validates a fractional factorial design in the marketing policy optimization space. This paper presents a framework that breaks down the business policy space into factors, accelerating learning and optimization velocity by improving statistical power given limited testable sample. Subsequently, we use a fractional factorial design to reduce the number of variants required to be implemented for experimentation, significantly reducing the implementation costs. Additionally, we continue to build on this methodology to leverage heterogeneous treatment effects and improve business outcomes by enabling optimal personalized policies. Furthermore, we have devised a robust evaluation procedure that facilitates the validation of model assumptions when dividing the policy space into factors.\\

In our business context, our framework enables us to discover a policy with 5\% incremental profit, with a 267\% higher experimentation speed and 67\% lower setup cost, relative to the status quo. This framework also presents a rare opportunity to run an HTE model on a randomized experimental sample. Exploiting the heterogeneity in treatment effects we can further improve the business impact by 2\%. Overall, we believe this framework can be applied to a broad category of experiments where the cardinality of the policy space is high and implementation costs are prohibitive. \\

\section{Acknowledgements}
\hfill \break
We are grateful to Elea Feit and Seenu Srinivasan for their comments and suggestions on this paper. We also express our gratitude to our partners at DoorDash, Kristin Mendez, Meghan Bender, Will Stone, and Taryn Riemer for helping us configure and launch the experiments and supporting us throughout this research. We also acknowledge the contributions of the data science and engineering community at DoorDash, especially Qiyun Pan, Caixia Huang, and Zhe Mai. Finally, we thank Jason Zheng, Bhawana Goel, and Sudhir Tonse. The completion of this research would not have been possible without your contribution and support. 

\hfill \break

\bibliographystyle{abbrv}
\bibliography{sigproc}  
%
%
\appendix
\section{HTE Interaction Test Results}
\hfill \break
Table \ref{table:1} shows results from regressing our business outcome on experimental factor-levels and their interactions with consumer characteristics. Overall, we do not detect significant systematic heterogeneity.\\

\begin{table}[!hbt]
\centering
\resizebox{\columnwidth}{3.1cm}{%
\begin{tabular}{lllllll} 
 \hline
   & Coefficient & Standard Error & Z & P-value & [0.025 & 0.975] \\ [0.5ex] 
 \hline\hline
Intercept                                                                     & 2.17916 & 0.48042 &  4.546   &  0.000  & 1.24030 & 3.11802 \\
C(promo\_spread){[}T.upfront{]}                                               & 0.40161 & 0.34854 &  1.156   &  0.248  & -0.27946 & 1.08330 \\
C(discount){[}Level 2{]}                                              & -0.37303 & 0.53066 &  -0.704  &  0.482  & -1.41300 & 0.66568 \\
C(discount){[}Level 3{]}                                          & 0.17239 & 0.42390 &  0.408   &  0.683  & -0.65626 & 1.00166 \\
C(triggering\_timing){[}T.weekday only{]}                                     & -0.02512 & 0.34854 &  -0.072  &  0.943  & -0.70650 & 0.65626 \\
C(messaging){[}T.mxrec{]}                                                     & 0.58435 & 0.34854 &  1.681   &  0.093  & -0.09734 & 1.26542 \\
avg-order-spend                                                                 & 0.21760 & 0.00942 &  21.420  &  0.000  & 0.19782 & 0.23864 \\
C(promo\_spread){[}T.upfront{]}:avg-order-spend                                 & -0.00973 & 0.00628 &  -1.256  &  0.209  & -0.02512 & 0.00628 \\
C(discount){[}Level 2{]}:avg-order-spend                                & 0.02386 & 0.01256 &  2.041   &  0.041  & 0.00000 & 0.04710 \\
C(discount){[}Level 3{]}:avg-order-spend                            & 0.01382 & 0.00942 &  1.606   &  0.108  & -0.00314 & 0.03140 \\
C(triggering\_timing){[}T.weekday only{]}:avg-order-spend                       & -0.00973 & 0.00628 &  -1.281  &  0.200  & -0.02512 & 0.00628 \\
C(messaging){[}T.mxrec{]}:avg-order-spend                                       & -0.00471 & 0.00628 &  -0.611  &  0.541  & -0.01884 & 0.00942 \\
resurrected\_dow                                                              & 0.13345 & 0.05024 &  2.707   &  0.007  & 0.03768 & 0.22922 \\
C(promo\_spread){[}T.upfront{]}:resurrected\_dow                              & 0.02041 & 0.03768 &  0.553   &  0.580  & -0.05338 & 0.09420 \\
C(discount){[}Level 2{]}:resurrected\_dow                             & -0.06939 & 0.05338 &  -1.302  &  0.193  & -0.17270 & 0.03454 \\
C(discount){[}Level 3{]}:resurrected\_dow                         & -0.07065 & 0.04396 &  -1.550  &  0.121  & -0.16014 & 0.01884 \\
C(triggering\_timing){[}T.weekday only{]}:resurrected\_dow                    & -0.01570 & 0.03768 &  -0.423  &  0.672  & -0.08792 & 0.05652 \\
C(messaging){[}T.mxrec{]}:resurrected\_dow                                    & 0.00126 & 0.03768 &  0.037   &  0.971  & -0.07222 & 0.07222 \\
churn\_tenure                                                                 & -0.00408 & 0.00000 &  -8.944  &  0.000  & -0.00628 & -0.00314 \\
C(promo\_spread){[}T.upfront{]}:churn\_tenure                                 & 0.00029 & 0.00031 &  0.920   &  0.358  & -0.00000 & 0.00000 \\
C(discount){[}Level 2{]}:churn\_tenure                                & 0.00031 & 0.00000 &  0.965   &  0.335  & -0.00000 & 0.00000 \\
C(discount){[}Level 3{]}:churn\_tenure                            & -0.00003 & 0.00000 &  -0.076  &  0.939  & -0.00000 & 0.00000 \\
C(triggering\_timing){[}T.weekday only{]}:churn\_tenure                       & 0.00010 & 0.00031 &  0.321   &  0.748  & -0.00000 & 0.00000 \\
C(messaging){[}T.mxrec{]}:churn\_tenure                                       & -0.00063 & 0.00031 &  -1.559  &  0.119  & -0.00000 & 0.00012 \\
promo\_lifetime\_orders                                                       & 0.13408 & 0.01256 &  11.266  &  0.000  & 0.10990 & 0.15700 \\
C(promo\_spread){[}T.upfront{]}:promo\_lifetime\_orders                       & 0.00785 & 0.00942 &  0.855   &  0.393  & -0.00942 & 0.02512 \\
C(discount){[}Level 2{]}:promo\_lifetime\_orders                      & 0.03799 & 0.01256 &  2.956   &  0.003  & 0.01256 & 0.06280 \\
C(discount){[}Level 3{]}:promo\_lifetime\_orders                  & 0.01790 & 0.00942 &  1.659   &  0.097  & -0.00314 & 0.04082 \\
C(triggering\_timing){[}T.weekday only{]}:promo\_lifetime\_orders             & 0.00314 & 0.00942 &  0.341   &  0.733  & -0.01570 & 0.02198 \\
C(messaging){[}T.mxrec{]}:promo\_lifetime\_orders                             & -0.00816 & 0.00942 &  -0.893  &  0.372  & -0.02512 & 0.00942 \\
orders\_prechurn30d                                                           & 1.04688 & 0.15386 &  6.736   &  0.000  & 0.74104 & 1.35020 \\
C(promo\_spread){[}T.upfront{]}:orders\_prechurn30d                           & -0.21446 & 0.11304 &  -1.915  &  0.056  & -0.43332 & 0.00628 \\
C(discount){[}Level 2{]}:orders\_prechurn30d                          & -0.41354 & 0.16642 &  -2.493  &  0.013  & -0.73790 & -0.08792 \\
C(discount){[}Level 3{]}:orders\_prechurn30d                      & -0.16768 & 0.14444 &  -1.156  &  0.248  & -0.45216 & 0.11618 \\
C(triggering\_timing){[}T.weekday only{]}:orders\_prechurn30d                 & 0.16579 & 0.11304 &  1.480   &  0.139  & -0.05338 & 0.38622 \\
C(messaging){[}T.mxrec{]}:orders\_prechurn30d                                 & -0.01068 & 0.11304 &  -0.094  &  0.925  & -0.22922 & 0.20724 \\
num\_visits\_during\_l30d\_ofchurn                                            & -0.31902 & 0.08164 &  -3.972  &  0.000  & -0.47728 & -0.16014 \\
C(promo\_spread){[}T.upfront{]}:num\_visits\_during\_l30d\_ofchurn            & -0.02229 & 0.05966 &  -0.371  &  0.711  & -0.14130 & 0.09734 \\
C(discount){[}Level 2{]}:num\_visits\_during\_l30d\_ofchurn           & -0.15135 & 0.08792 &  -1.734  &  0.083  & -0.32342 & 0.01884 \\
C(discount){[}Level 3{]}:num\_visits\_during\_l30d\_ofchurn       & -0.08886 & 0.07222 &  -1.222  &  0.222  & -0.23236 & 0.05338 \\
C(triggering\_timing){[}T.weekday only{]}:num\_visits\_during\_l30d\_ofchurn  & -0.05087 & 0.05966 &  -0.842  &  0.400  & -0.16956 & 0.06908 \\
C(messaging){[}T.mxrec{]}:num\_visits\_during\_l30d\_ofchurn                  & -0.18338 & 0.05966 &  -3.030  &  0.002  & -0.30144 & -0.06594 \\
num\_email\_during\_l30d\_ofchurn                                             & -0.06782 & 0.01884 &  -3.655  &  0.000  & -0.10362 & -0.03140 \\
C(promo\_spread){[}T.upfront{]}:num\_email\_during\_l30d\_ofchurn             & 0.01225 & 0.01570 &  0.842   &  0.400  & -0.01570 & 0.04082 \\
C(discount){[}Level 2{]}:num\_email\_during\_l30d\_ofchurn            & -0.01947 & 0.01884 &  -0.964  &  0.335  & -0.05966 & 0.01884 \\
C(discount){[}Level 3{]}:num\_email\_during\_l30d\_ofchurn        & -0.01005 & 0.01570 &  -0.583  &  0.560  & -0.04396 & 0.02198 \\
C(triggering\_timing){[}T.weekday only{]}:num\_email\_during\_l30d\_ofchurn   & 0.01099 & 0.01570 &  0.771   &  0.441  & -0.01570 & 0.03768 \\
C(messaging){[}T.mxrec{]}:num\_email\_during\_l30d\_ofchurn                   & -0.00440 & 0.01570 &  -0.311  &  0.756  & -0.03140 & 0.02512 \\
num\_push\_during\_l30d\_ofchurn                                              & -0.05118 & 0.00942 &  -4.988  &  0.000  & -0.07222 & -0.03140 \\
C(promo\_spread){[}T.upfront{]}:num\_push\_during\_l30d\_ofchurn              & 0.00220 & 0.00942 &  0.265   &  0.791  & -0.01256 & 0.01884 \\
C(discount){[}Level 2{]}:num\_push\_during\_l30d\_ofchurn             & 0.00220 & 0.01256 &  0.182   &  0.856  & -0.01884 & 0.02512 \\
C(discount){[}Level 3{]}:num\_push\_during\_l30d\_ofchurn         & -0.00597 & 0.00942 &  -0.652  &  0.515  & -0.02512 & 0.01256 \\
C(triggering\_timing){[}T.weekday only{]}:num\_push\_during\_l30d\_ofchurn    & -0.00879 & 0.00942 &  -1.079  &  0.281  & -0.02512 & 0.00628 \\
C(messaging){[}T.mxrec{]}:num\_push\_during\_l30d\_ofchurn                    & 0.01005 & 0.00942 &  1.237   &  0.216  & -0.00628 & 0.02512 \\
\end{tabular}%
}
\caption{HTE Interaction Test Results}
\label{table:1}
\end{table}

\pagebreak

\begin{table}
\centering
\resizebox{\columnwidth}{3.1cm}{
\begin{tabular}{cccccc}
\toprule
Promo\_spread & Discount & Triggering timing & Messaging &  Predicted Profit & In sample \\
\midrule
Upfront  &  Level3  &  Ongoing  &  Generic  & 7.90091 &  False \\

Upfront  &  Level2  &  Ongoing  &  Generic  & 7.88897 &  True \\

Upfront  &  Level3  &  weekday only  &  Generic  & 7.88752 &  False \\

Upfront  &  Level2  &  weekday only  &  Generic  & 7.87558 &  True \\

Upfront  &  Level3  &  Ongoing  &  Mxrec  & 7.87374 &  True \\

Upfront  &  Level2  &  Ongoing  &  Mxrec  & 7.86181 &  False \\

Upfront  &  Level3  &  weekday only  &  Mxrec  & 7.86036 &  False \\

Upfront  &  Level2  &  weekday only  &  Mxrec  & 7.84842 &  False \\

Spread  &  Level3  &  Ongoing  &  Generic  & 7.77188 &  False \\

Spread  &  Level2  &  Ongoing  &  Generic  & 7.75994 &  False \\

Spread  &  Level3  &  weekday only  &  Generic  & 7.75849 &  True \\

Spread  &  Level2  &  weekday only  &  Generic  & 7.74656 &  False \\

Spread  &  Level3  &  Ongoing  &  Mxrec  & 7.74472 &  False \\

Spread  &  Level2  &  Ongoing  &  Mxrec  & 7.73278 &  True \\

Spread  &  Level3  &  weekday only  &  Mxrec  & 7.73133 &  False \\

Spread  &  Level2  &  weekday only  &  Mxrec  & 7.71939 &  True \\

Upfront  &  Level1  &  Ongoing  &  Generic  & 7.48146 &  False \\

Upfront  &  Level1  &  weekday only  &  Generic  & 7.46807 &  False \\

Upfront  &  Level1  &  Ongoing  &  Mxrec  & 7.45429 &  False \\

Upfront  &  Level1  &  weekday only  &  Mxrec  & 7.44091 &  True \\

Spread  &  Level1  &  Ongoing  &  Generic  & 7.35242 &  True \\

Spread  &  Level1  &  weekday only  &  Generic  & 7.33904 &  False \\

Spread  &  Level1  &  Ongoing  &  Mxrec  & 7.32526 &  False \\

Spread  &  Level1  &  weekday only  &  Mxrec  & 7.31188 &  False \\
\bottomrule
\end{tabular}
}
\caption{Predicted Profit from All policies}
\label{tab:policypred}
\end{table}

\end{document}